\documentclass[aps,prl,amsmath,amssymb,floatfix,twocolumn,amsmath,superscriptaddress,twocolumn,nofootinbib,tighten,letterpaper]{revtex4-2}

\usepackage{tikz}
\usepackage{lmodern}

\usepackage{amsbsy}
\usepackage{amsthm}
\usepackage[autostyle]{csquotes}
\usepackage{setspace}
\usepackage{svg}
\usepackage{bm}
\usepackage{placeins}

\usepackage{tabularray}
\definecolor{RowColor}{rgb}{0.88,1,0.9}
\newcommand\myeq{\mkern1.5mu{=}\mkern1.5mu}

\usepackage{textgreek}

\usepackage{multirow}
\usepackage{subfigure}
\usepackage{color}
\usepackage{mathrsfs}
\usepackage{hyperref}
\usepackage[normalem]{ulem}
\usepackage{bm}

\usepackage{amssymb}   
\usepackage{amsmath}
\renewcommand\vec[1]{\ensuremath\boldsymbol{#1}} 

\usepackage{amsfonts, relsize}
\usepackage{graphics}
\usepackage{graphicx}
\usepackage{subfigure}
\usepackage{hyperref}
\usepackage{color}
\usepackage{comment}

\newcommand{\brac}[1]{\left[ #1 \right]}

\begin{document}

\title{Polar hairs of mixed-parity nodal superconductors in Rarita-Schwinger-Weyl metals}

\author{Saswata Mandal}
\affiliation{Department of Physics, Indian Institute of Science, Bangalore 560012, India}

\author{Bitan Roy}
\affiliation{Department of Physics, Lehigh University, Bethlehem, Pennsylvania, 18015, USA}

\date{\today}

\begin{abstract}
Linearly dispersing Rarita-Schwinger-Weyl (RSW) fermions featuring two Fermi velocities are the key constituents of itinerant spin-3/2 quantum materials. When doped, RSW metals sustain two Fermi surfaces (FSs), around which one fully gapped $s$-wave and five \emph{mixed-parity} local pairings can take place. The intraband components of four mixed-parity pairings support point nodes at the poles of two FSs, only around which long-lived quasiparticles live. For weak (strong) pairing amplitudes ($\Delta$), gapless north and south poles belonging to the same (different) FS(s) get connected by \emph{polar hairs}, one-dimensional line nodes occupying the region between two FSs. The remaining one, by contrast, supports four nodal rings in between two FSs, symmetrically placed about their equators, but only when $\Delta$ is small. For large $\Delta$, this paired state becomes fully gapped. The transition temperature and pairing amplitudes follow the BCS scaling. We explicitly showcase these outcomes for a rotationally symmetric RSW metal, and contrast our findings when the system possesses an enlarged Lorentz symmetry and with those in spin-3/2 Luttinger materials.      
\end{abstract}

\maketitle

\emph{Introduction}.~Emergent phenomena lead to peculiar outcomes in quantum solids~\cite{anderson}. Electron and hole quasiparticle excitations, respectively characterized by positive and negative effective mass and charge, stand as its paradigmatic examples~\cite{ashcroftmarmin}. More intriguingly, spin-1/2 electrons hopping under specific crystal environments can give rise to effective spin-3/2 excitations, as is the case in 227 pyrochlore iridates~\cite{spin32:lutt1, spin32:lutt2, spin32:lutt3}, half-Heuslers~\cite{spin32:lutt4, spin32:lutt5}, HgTe~\cite{spin32:lutt6} and gray tin~\cite{spin32:lutt7, spin32:lutt8}. In these materials spin-3/2 Luttinger fermions display a biquadratic touching between Kramers degenerate valence and conduction bands, with respective spin projections $j=\pm 3/2$ and $\pm 1/2$~\cite{luttinger}. Spin-3/2 Rarita-Schwinger-Weyl (RSW) fermions~\cite{RSW}, by contrast, display linear band touching, featuring two Fermi velocities $\pm v/2$ and $\pm 3 v/2$. In RSW materials the valence (conduction) band is composed of $j=-3/2,-1/2$ ($3/2,1/2$) spin projections. While antiperovskites harbor massive RSW fermions~\cite{spin32:met1}, PdGa~\cite{spin32:met2}, PdBiSe~\cite{spin32:met3}, AlPt~\cite{spin32:met4}, PtGa~\cite{spin32:met5}, CoSi~\cite{spin32:met6, spin32:met7, spin32:met8} and RhSi~\cite{spin32:met6} accommodate their gapless counterpart. Altogether, these electronic materials create an exciting opportunity to study exotic Cooper pairings among spin-3/2 fermions. While the Cooper avenue has been explored in Luttinger materials~\cite{LuttingerSC:th0, LuttingerSC:th1, LuttingerSC:th2, LuttingerSC:th3, LuttingerSC:th4, LuttingerSC:th5, LuttingerSC:th6, LuttingerSC:th7, LuttingerSC:th8, LuttingerSC:th9, LuttingerSC:th10, LuttingerSC:th11, LuttingerSC:th12}, due to experimental pertinence in superconducting half-Heuslers~\cite{LuttingerSC:exp1, LuttingerSC:exp2, LuttingerSC:exp3, LuttingerSC:exp4}, pairing among RSW fermions is still in its infancy~\cite{RSW:th1, RSW:th2, RSW:th3, RSW:th4, RSW:th5}.

\begin{figure}[t!]
\includegraphics[width=0.95\linewidth]{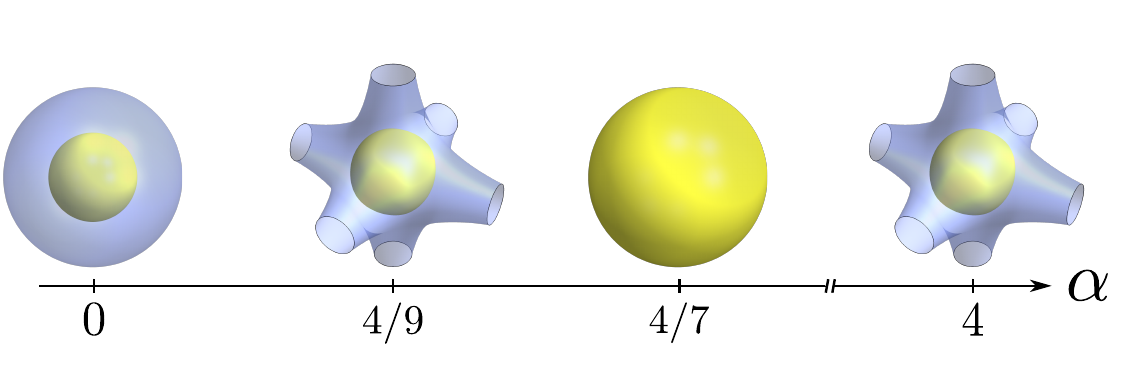}
\caption{Topology of Fermi surfaces for various $\alpha$. See Eq.~(\ref{eq:RSWHamil}).    
}~\label{fig:fermisurfacephasediagram}
\end{figure}

\begin{figure*}[t!]
\includegraphics[width=0.95\linewidth]{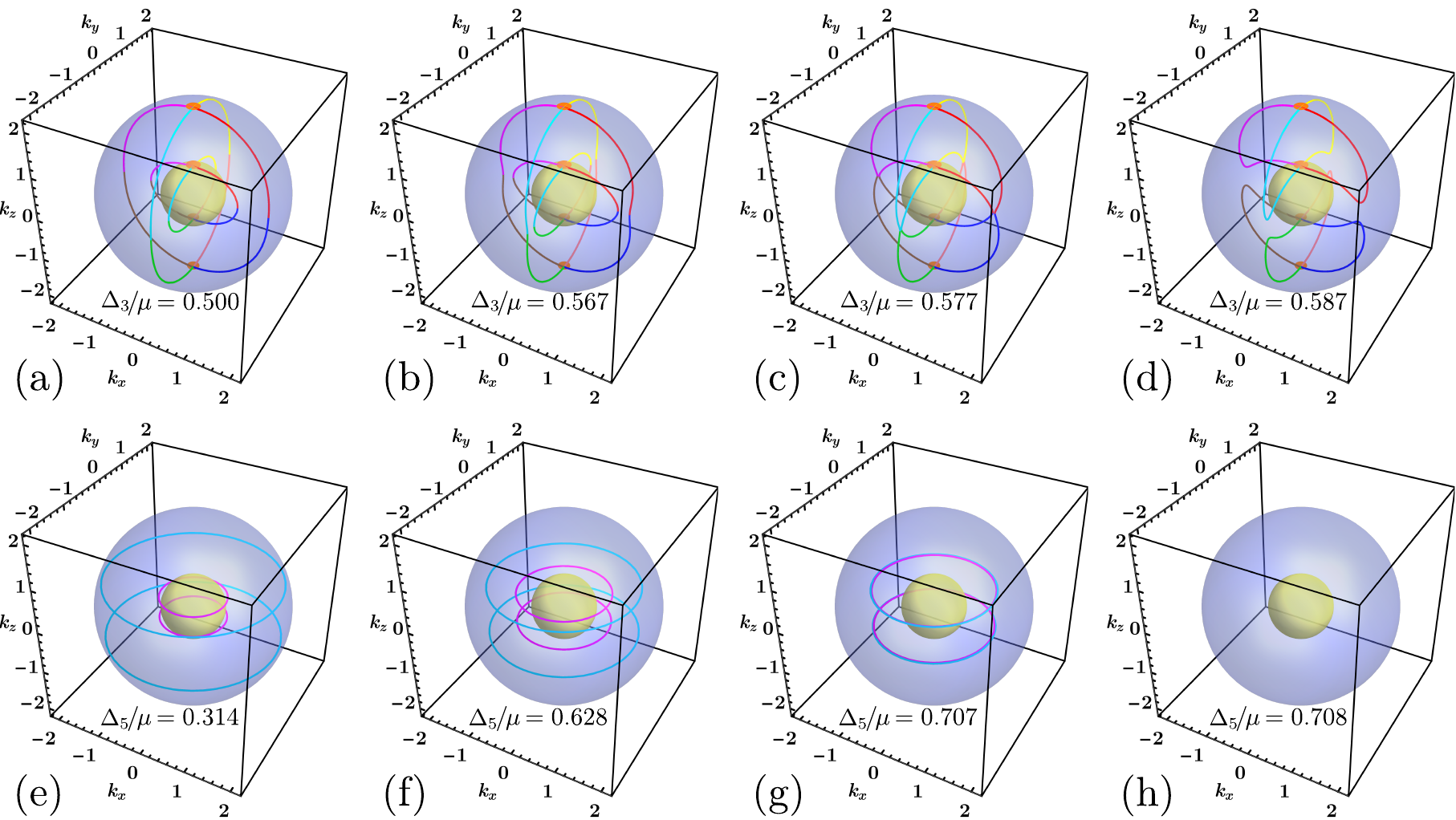}
\caption{Evolution of polar hairs (top) and nodal rings (bottom) with pairing amplitude ($\Delta_j/\mu$) for $\alpha=0$. Polar hairs (nodal rings) are shown for $\Delta_{3}$ ($\Delta_5$) pairing. (a),(b) Different segments of polar hairs (color coded) connecting the north and south poles (red dots) of the same FS, (c) touch each other at a critical amplitude, (d) beyond which they connect same poles of different FSs. (e),(f) Nodal rings, symmetrically placed about the equators, get closer with increasing $\Delta_5$. (g) They touch each other for a critical amplitude. (h) The paired state then becomes fully gapped. Momentum $\vec{k}$ is measured in units of $\mu/v$.    
}~\label{fig:nodalloopevolution}
\end{figure*}

\begin{figure}[b!]
\includegraphics[width=0.90\linewidth]{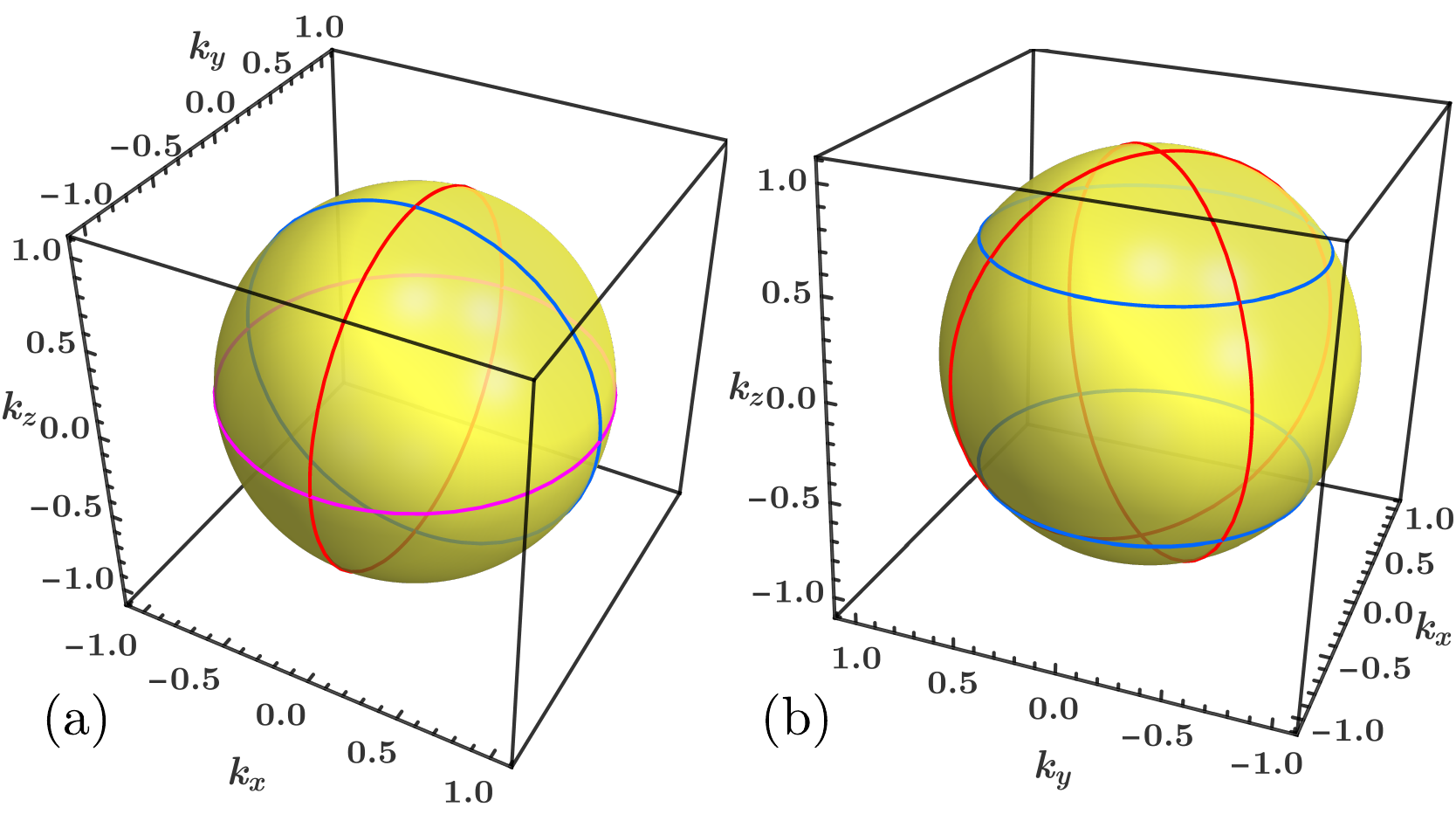}
\caption{(a) Nodal loops for $\Delta_1$ (red), $\Delta_2$ (blue) and $\Delta_3$ (magenta) pairings in a Lorentz symmetric RSW metal with a single FS of Fermi momentum $k_{F}=7\mu/(3v)$. We set $\mu/v=3/7$. (b) Nodal loops for $\Delta_4$ (red) and $\Delta_5$ (blue) pairings in a Luttinger metal. Here momenta $\vec{k}$ are measured in units of $k_F=\sqrt{2 m_\pm \mu}$. Nodal loops for $\Delta_1$, $\Delta_2$, and $\Delta_3$ pairings are obtained from those for $\Delta_4$ pairing after suitable rotation~\cite{LuttingerSC:th7}.       
}~\label{fig:nodalloopLorentz}
\end{figure}

\emph{key results}.~In crystals, RSW fermions are accompanied by a doubler (generalized Nielsen-Ninomiya theorem~\cite{nielsenninomiya, SCzhangdoubling}). Here, for simplicity we consider a single irreducible copy of four-component RSW fermions. Such gapless excitations, nonetheless, can be found on the three-dimensional hypersurface of a four-dimensional bulk topological insulator of massive spin-3/2 fermions~\cite{kennetetal:spin32}. When doped, RSW metals typically sustain two Fermi surfaces (FSs) (Fig.~\ref{fig:fermisurfacephasediagram}). Around them exotic local \emph{mixed-parity} nodal superconductors can nucleate (Table~\ref{tab:pairingsymmetry}). Their intraband components often feature gapless \emph{polar hairs} (Fig.~\ref{fig:nodalloopevolution}), one-dimensional line nodes connecting the poles of the FSs through its interior, in the presence of an effective attraction within a characteristic Debye frequency (weak-coupling BCS mechanism). These outcomes are contrasted with those when the RSW materials enjoy enlarged Lorentz symmetry and in Luttinger materials, featuring a single FS. In both cases, nodal pairings host simple line nodes on the FS (Fig.~\ref{fig:nodalloopLorentz}).   

\emph{Model}.~The Hamiltonian describing RSW fermions is
\begin{equation}~\label{eq:RSWHamil}
H_{\rm RSW}= v \left( \vec{J} \cdot \vec{k} - \alpha \; \vec{J}^3 \cdot \vec{k} \right)-\mu,
\end{equation}
where $v$ bears the dimension of the Fermi velocity~\cite{RSW:th2, RSW:th3, RSW:th4, RSW:topo1, RSW:topo2}. Momentum $\vec{k}$ and chemical potential $\mu$ are measured from the band touching RSW node, $\vec{J}=(J_x, J_y, J_z)$ is the vector spin-3/2 matrix with $J_z={\rm diag}. (3,1,-1,-3)/2$, $\vec{J}^3=(J^3_x,J^3_y, J^3_z)$, and $\alpha$ is a tuning parameter. Spin-3/2 matrices satisfy the SU(2) algebra $[J_p,J_q]=i \epsilon_{pqr} J_r$, where $p,q,r=x,y,z$ and $\epsilon_{pqr}$ is the antisymmetric tensor. For $\alpha=0$ the system possesses a rotational symmetry, yielding two isotropic FSs with Fermi momenta $k_{F,j}=\mu/(j v)$ with $j=1/2,3/2$. By contrast, a Lorentz symmetry emerges when $\alpha=4/7$ and the system supports a doubly degenerate FS with Fermi momentum $k_F=7\mu/(3v)$. Then matrices appearing in $H_{\rm RSW}$, $\vec{J}-4 \vec{J}^3/7= 3\{ \Gamma_{23}, \Gamma_{31}, \Gamma_{12} \}/7$ besides the SU(2) algebra also satisfy the mutually anticommuting Clifford algebra. Here $\Gamma_{jk}=[\Gamma_j,\Gamma_k]/(2i)$ and five mutually anticommuting Hermitian matrices are $\Gamma_1=\kappa_3 \sigma_2$, $\Gamma_2=\kappa_3 \sigma_1$, $\Gamma_3=\kappa_2 \sigma_0$, $\Gamma_4=\kappa_1 \sigma_0$ and $\Gamma_5=\kappa_3 \sigma_3$. Two sets of Pauli matrices $\{ \kappa_\nu\}$ and $\{ \sigma_\nu \}$ operate on the sign ($\pm$) and magnitude ($3/2,1/2$) of the spin projections, respectively~\cite{spin3/2Gamma:1, spin3/2Gamma:2}. Nodal topology of RSW fermions changes at $\alpha=4/9$ and $4$~\cite{RSW:th3, RSW:topo1, RSW:topo2} at which two FSs touch along the principal axes [Fig.~\ref{fig:fermisurfacephasediagram}].

\begin{table*}[t!]
\begin{tblr}{width = \linewidth, colspec = {|Q[c,m,1.1cm]|[2pt,white]Q[l,m,3cm]|[2pt,white]Q[c,m,0.93cm]|[2pt,white]Q[c,m,2.66cm]|[2pt,white]Q[l,m,3cm]|[2pt,white]Q[c,m,0.93cm]|[2pt,white]Q[c,m,2.cm]|},
cell{odd}{1-7} = {RowColor},
row{1-3} = {white},}
\hline
 \SetCell[r=3]{c}\small{\textbf{Pairing\\matrix\\\footnotesize{(Ampli-tude)}}} 
 & \SetCell[c=6]{c} \textbf{Pairing near Fermi surfaces and emergent topology in RSW metals} \\
 \hline
    &\SetCell[c=3]{c} \textbf{$\bm{\alpha=0}$ (Rotational symmetry)}  &&&\SetCell[c=3]{c}\textbf{{$\bm{\alpha=4/7}$} (Lorentz symmetry)}\\
\cmidrule[lr]{2-4}
\cmidrule[lr]{5-7}
&\small{\textbf{Angular functions}}& \small{\textbf{Parity}}&\small{\textbf{Nodal lines}} &\small{\textbf{Angular functions}}& \small{\textbf{Parity}}&\small{\textbf{Nodal lines}}   \\
 \hline
 \SetCell[r=1]{c}$\Gamma_0(\Delta_0)$&$a_0^0\myeq1$&+&$\times$&$a_0^0\myeq1$&+&$\times$\\
{\\$\Gamma_1(\Delta_1)$\\}&{\small{$a_1^1\myeq\cos(2\theta)\sin(\phi)$ \\[1mm]$a_2^1\myeq\cos(\theta)\cos(\phi)$\\[1mm]$a_3^1\myeq\frac{\sqrt{3}}{2}\sin(2\theta)\sin(\phi)$}}&{--\\+\\+}&{\footnotesize{$q_{y}\myeq 0,\delta^2_1 q_z^2 \myeq f(q)q^2$\\[2mm] $q_{z}\myeq 0,\delta^2_1 q_y^2 \myeq f(q) q^2$}}&\small{$a_1^1 \myeq \frac{1}{2}\sin(\theta)\sin(2\phi)$\\[2mm]$a_2^1\myeq\sin(\theta)\cos^2(\phi)$}&{+\\[2mm]+}&\small{$q_x\myeq0$,\\[2mm]$q_y^2+q_z^2\myeq\left(\frac{7}{3}\right)^2$}\\
 {\\$\Gamma_2(\Delta_2)$\\}&{\small{$a_1^2\myeq\cos(2\theta)\cos(\phi)$ \\[1mm]$a_2^2\myeq\cos(\theta)\sin(\phi)$\\[1mm]$a_3^2\myeq\frac{\sqrt{3}}{2}\sin(2\theta)\cos(\phi)$}}&{--\\+\\+}&{\footnotesize{$q_{z}\myeq 0,\delta^2_2 q_x^2 \myeq f(q) q^2$\\[2mm] $q_{x}\myeq 0,\delta^2_2 q_z^2 \myeq f(q) q^2$}}&\small{$a_1^2 \myeq \sin(\theta)\sin^2(\phi)$\\[2mm]$a_2^2\myeq\frac{1}{2}\sin(\theta)\sin(2\phi)$}&{+\\[2mm]+}&\small{$q_y\myeq0$,\\[2mm]$q_z^2+q_x^2\myeq\left(\frac{7}{3}\right)^2$}\\
 {\\$\Gamma_3(\Delta_3)$\\}&{\small{$a_1^3\myeq\frac{1}{2}\sin(2\theta)\sin(2\phi)$ \\[1mm]$a_2^3\myeq\sin(\theta)\cos(2\phi)$\\[1mm]$a_3^3\myeq\frac{\sqrt{3}}{2}\sin^2(\theta)\sin(2\phi)$}}&{--\\+\\+}&{\footnotesize{$q_{x}\myeq 0,\delta^2_3 q_y^2 \myeq f(q)q^2$\\[2mm] $q_{y}\myeq 0,\delta^2_3 q_x^2 \myeq f(q)q^2$}}&\small{$a_1^3 \myeq \cos(\theta)\sin(\phi)$\\[2mm]$a_2^3\myeq\cos(\theta)\cos(\phi)$}&{+\\[2mm]+}&\small{$q_z\myeq0$,\\[2mm]$q_x^2+q_y^2\myeq\left(\frac{7}{3}\right)^2$}\\
{\\$\Gamma_4(\Delta_4)$\\}&{\small{$a_1^4\myeq\frac{1}{2}\sin(2\theta)\cos(2\phi)$ \\[1mm]$a_2^4\myeq\sin(\theta)\sin(2\phi)$\\[1mm]$a_3^4\myeq\frac{\sqrt{3}}{2}\sin^2(\theta)\cos(2\phi)$}}&{--\\+\\+}&{\footnotesize{$q_{x}\myeq \pm q_y$\\[2mm] $\delta^2_4 q_z^2 \myeq [\delta^2_4-f(q)]q^2$}}&\small{$a_1^4 \myeq\cos(\phi)$\\[2mm]$a_2^4\myeq\sin(\phi)$}&{--\\[2mm]--}&\small{$\times$}\\ 
{\\$\Gamma_5(\Delta_5)$\\}&{\small{$a_1^5\myeq\frac{\sqrt{3}}{2}\sin(2\theta)$\\[1mm]$a_3^5\myeq\frac{1}{4}(3\cos^2(\phi)+1)$}}&{--\\+}&{\footnotesize{$q_x^2+q_y^2\myeq \frac{8}{9}q_{\pm}^2$\\[1mm] $q_z^2 \myeq \pm\frac{1}{9}q_{\pm}^2$}}&\small{$a_3^5 \myeq 1$}&{+}&\small{$\times$}\\ 
\hline
\end{tblr}
\caption{Symmetry classification and emergent nodal topology of six local pairings appearing in Eqs.~(\ref{eq:BdGfull}) and (\ref{eq:reducedBCS}) for RSW fermions around the FSs for $\alpha=0$ and $4/7$. The even-parity $s$-wave pairing ($\Delta_0$) is always fully gapped. The $\Delta_1$, $\Delta_2$, $\Delta_3$, and $\Delta_4$ pairings are always gapless when $\alpha=0$. By contrast, the $\Delta_5$ pairing is gapless up to a critical pairing amplitude $\delta^\ast_5=1/\sqrt{2}$, beyond which it becomes fully gapped. See Fig.~\ref{fig:nodalloopevolution}. These five non-$s$-wave pairings are mixed-parity in nature for $\alpha=0$. Here $+$ ($-$) corresponds to even (odd) under the parity ($\theta \to \pi-\theta$ and $\phi \to \pi + \phi$). By contrast, $\Delta_4$ ($\Delta_1$, $\Delta_2$, $\Delta_3$, and $\Delta_5$) pairing(s) is (are) odd (even) under parity when $\alpha=4/7$. Then only $\Delta_1$, $\Delta_2$, and $\Delta_3$ pairings are gapless. See Fig.~\ref{fig:nodalloopLorentz}(a). The equations for the polar hairs and nodal rings are in terms of dimensionless momenta $q_j=v k_j/\mu$ and pairing amplitude $\delta_\nu=\Delta_\nu/\mu$, where $q^2=q^2_x+q^2_y+q^2_z$, $f(q)= [q^2-4 (q-1)^2]/4$, and $q_\pm =2 [2 \pm (1-2 \delta^2)^{1/2}]/3$. For spin-3/2 Luttinger fermions (displaying biquadratic band touching in the normal state), the angular dependence of $\Delta_\nu$ pairings is solely captured by $a^\nu_3$ for $\alpha=0$, which can be expressed in terms of cubic $d$-wave harmonics (even under parity). Each of them supports two nodal loops [Fig.~\ref{fig:nodalloopLorentz}(b)]. Missing $a^\nu_\rho$s are trivial.     
}~\label{tab:pairingsymmetry}
\end{table*}

\emph{Nambu doubling}.~To capture superconductivity in RSW materials, we Nambu double $H_{\rm RSW}$. After absorbing a unitary matrix $\Gamma_{13}$ in the hole part of the Nambu spinor, the normal-state Hamiltonian reads $H^{\rm Nam}_{\rm RSW}=\eta_3 H_{\rm RSW}$. Newly introduced Pauli matrices $\{ \eta_\nu \}$ operate on the Nambu indices. The antisymmetric nature of the effective single-particle Bogoliubov de Gennes (BdG) Hamiltonian, manifesting the Pauli exclusion principle~\cite{comment:1}, in the presence of all the local or on-site or momentum-independent pairings restricts its form to 
\allowdisplaybreaks[4]
\begin{equation}~\label{eq:BdGfull}
H_{\rm BdG} = \left( \eta_1 \cos \phi + \eta_2 \sin \phi \right) \sum^{5}_{\nu=0} \Delta_\nu \Gamma_\nu,
\end{equation}   
where $\Gamma_0=\kappa_0\sigma_0$ and $\Delta_\nu$ are the pairing amplitudes for $\nu=0, \cdots, 5$. Without any loss of generality, we set the U(1) superconducting phase $\phi=0$. To unveil the emergent topology of the pairings near the FSs, next we project $H_{\rm BdG}$ onto the valence or conduction bands and consider only their \emph{intraband} pieces. This procedure is justified within the framework of weak-coupling BCS picture, in which attractive pairing interaction persists only around the FSs within a shell set by the Debye frequency. The reduced BCS Hamiltonian then reads 
\allowdisplaybreaks[4]
\begin{equation}~\label{eq:reducedBCS}
\hspace{-0.25cm} H^{\rm RSW}_{\rm BCS}= \eta_3 \left[ v|\vec{k}| \hat{h}_0 (\alpha) -\mu \sigma_0 \right] + \eta_1 \sum^{5}_{\nu=1} \Delta_\nu  \sum^3_{\rho=0} \sigma_\rho a^\nu_\rho,   
\end{equation}
where $\hat{h}_0 (0)={\rm diag}.(3,1)/2$ and $\hat{h}_0 (4/7)=(3/7) \sigma_0$. The angular dependence of the pairing terms $a^\nu_\rho \equiv a^\nu_\rho (\hat{\Omega},\alpha)$ for $\alpha=0$ and $4/7$ are shown in Table~\ref{tab:pairingsymmetry}, where $\hat{\Omega} \equiv (\theta, \phi)$ with $\theta \; (\phi)$ as the polar (azimuthal) angle. A detailed derivation is shown in the Supplemental Materials~\cite{supplementary}.

\emph{Parity}.~The angular functions $a^\nu_\rho$ allow us to pin the parity (${\mathcal P}$) of each paired state, under which $\theta \to \pi-\theta$, $\phi \to \pi+\phi$ [Table~\ref{tab:pairingsymmetry}]. For any $\alpha$, the $s$-wave $\Delta_0$ pairing is even under ${\mathcal P}$. The remaining five paired states correspond to mixed-parity pairings when $\alpha=0$, as $a^\nu_1$ ($a^\nu_{2,3}$) is (are) odd (even) under parity for $\nu=1,\cdots, 5$. For $\alpha=4/7$, parity eigenstates fragment into two sectors, and $\Delta_4$ ($\Delta_1$, $\Delta_2$, $\Delta_3$, and $\Delta_5$) pairing(s) is (are) even (odd) under ${\mathcal P}$. The parity mixing or fragmentation of the paired states \emph{solely} stems from the linear band dispersion of RSW quasiparticles, captured by $\vec{k}$-linear terms [Eq.~(\ref{eq:RSWHamil})] that are odd under ${\mathcal P}$. Therefore, paired states inherit parity from the normal state~\cite{pairingsymmnormalst:1, pairingsymmnormalst:2}, which we further justify from their even-parity counterparts in Luttinger materials, where the normal-state Hamiltonian is described in terms of even-parity $d$-wave harmonics [Eq.~\eqref{eq:luttingerhamil}] and consequently all the six local pairings are even under parity ($s$-wave or $d$-wave) [Eq.~\eqref{eq:luttingerpairing}].   

\begin{figure*}[t!]
\includegraphics[width=1.00\linewidth]{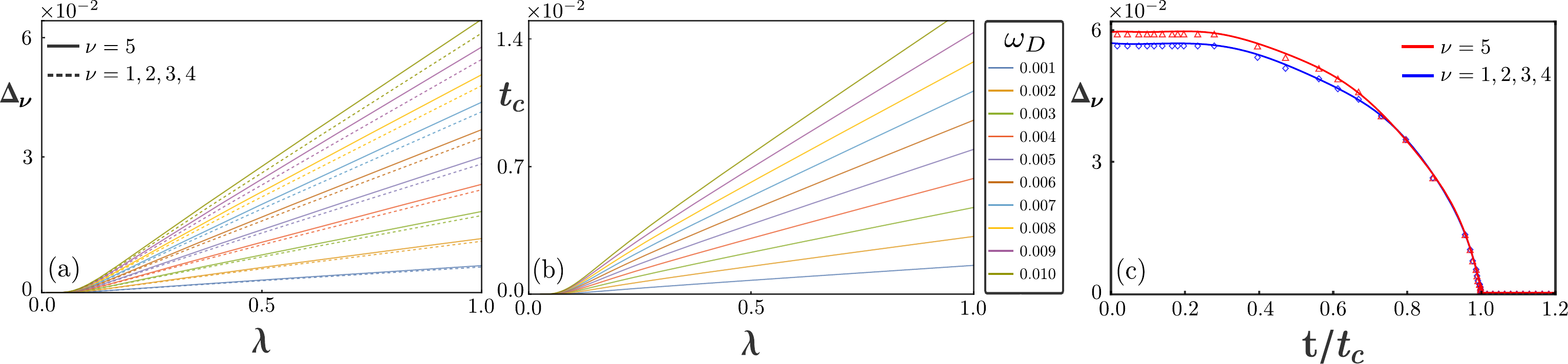}
\caption{Self-consistent solutions of (a) pairing amplitude ($\Delta_\nu$) and (b) transition temperature ($t_c$) with the pairing interaction strength ($\lambda$) for various Debye frequencies ($\omega_D$), when $\alpha=0$. Here, $\Delta_\nu$, $t_c$, $\lambda$, and $\omega_D$ are dimensionless (see text for definitions). (c) Scaling of $\Delta_\nu$ with temperature ($t/t_c$) for $\lambda=1$ and $\omega_D=0.01$, showing that despite possessing different amplitudes for $t<t_c$, all the pairing channels have \emph{identical} transition temperature. Results match the BCS scaling theory [Eq.~(\ref{eq:BCSscaling})].          
}~\label{fig:BCStheory}
\end{figure*}

\emph{BdG bands}.~The energy spectra $\pm E^\nu_j (\alpha)$ of $H^{\rm RSW}_{\rm BCS}$ are 
\allowdisplaybreaks[4]
\begin{eqnarray}
E^\nu_j (0) &=& \bigg[ \Delta^2_\nu  [(a^\nu_0)^2 + (a^\nu_3)^2 ] + \bigg(  (-1)^{j+1/2} v |\vec{k}|/2 \nonumber \\
&+& \sqrt{(v |\vec{k}|-\mu)^2 + \Delta^2_\nu [ (a^\nu_1)^2 + (a^\nu_2)^2]} \bigg)^2 \bigg]^{1/2}, \nonumber \\
E^\nu_j (4/7) &=& \bigg[ \left( \frac{3}{7} v |\vec{k}|-\mu \right)^2 + \Delta^2_\nu \sum^3_{\rho=0} (a^\nu_\rho)^2 \bigg]^{1/2}
\end{eqnarray}
for $j=1/2,3/2$. The independence of $E^\nu_j (4/7)$ on $j$ confirms the double degeneracy of each eigenvalue. Emergent nodal topology inside the paired states is computed from the zeros of $E^\nu_j (\alpha)$, which we discuss next.

\emph{Nodal topology}.~The even-parity $\Delta_0$ pairing always represents a fully gapped trivial $s$-wave superconductor. For $\alpha=0$, the $\Delta_1$, $\Delta_2$, $\Delta_3$ and $\Delta_4$ pairings are always gapless and support polar hairs, one-dimensional line nodes connecting the opposite (same) poles belonging to the same (different) FS(s) below (above) a critical amplitude $\Delta_j/\mu=1/\sqrt{3}$~\cite{supplementary}. They interpolate through the region between two FSs. The evolution of the polar hairs for the $\Delta_3$ pairing is shown in Fig.~\ref{fig:nodalloopevolution}(top) and its analytical equations are given in Table~\ref{tab:pairingsymmetry}. The structures of the polar hairs for $\Delta_1$ ($\Delta_2$) and $\Delta_4$ pairings can be obtained from those for the $\Delta_3$ pairing after $\pi/2$ ($\pi/2$) and $\pi/4$ rotations about $x$ ($y$) and $z$ axes, respectively. The $\Delta_5$ pairing, on the other hand, supports four nodal rings, symmetrically placed about the equators of the FSs and in the region in between them, but only up to a critical amplitude $\Delta_5/\mu=1/\sqrt{2}$ [Fig.~\ref{fig:nodalloopevolution}(bottom)]. Above this threshold amplitude, the $\Delta_5$ pairing is fully gapped. For $\alpha=4/7$, two nodal loops residing on a doubly degenerate FS for $\Delta_1$, $\Delta_2$, and $\Delta_3$ pairings can be rotated into each other [Fig.~\ref{fig:nodalloopLorentz}(a)]. By contrast, $\Delta_4$ and $\Delta_5$ pairings are then fully gapped. See Table~\ref{tab:pairingsymmetry}.

\emph{Luttinger Metal}.~The normal-state Hamiltonian for isotropic spin-3/2 Luttinger materials reads as~\cite{luttinger} 
\begin{equation}~\label{eq:luttingerhamil}
H_{\rm Lutt}=\left( \frac{\vec{k}^2}{2 m_0}-\mu \right) \Gamma_0 -\frac{\vec{k}^2}{2 m} \sum^5_{j=1} \Gamma_j \hat{d}_j(\hat{\Omega}). 
\end{equation}
Here $m_0$ and $m$ bear the dimension of mass, and $\hat{d}_j(\hat{\Omega})$ is a five-dimensional unit vector that transforms in the $l=2$ (``$d$-wave") representation under orbital SO(3) rotations. Its components are given by the cubic harmonics, linear combinations of the spherical harmonics $Y^m_{l=2}(\hat{\Omega})$~\cite{supplementary}. The BdG Hamiltonian for local pairings in this system is also given by $H_{\rm BdG}$ [Eq.~(\ref{eq:BdGfull})]. After the projection onto the valence ($-$) or conduction ($+$) band, the reduced BCS Hamiltonian takes the form 
\begin{equation}~\label{eq:luttingerpairing}
H^{\rm Lutt}_{\rm BCS}=\bigg[ \pm \frac{\vec{k}^2}{2 m_\pm} -\mu \bigg] \eta_3 + \eta_1 \bigg[ \Delta_0 + \sum^5_{j=1} \Delta_j a^{j}_3 (\hat{\Omega}, 0) \bigg],
\end{equation} 
where $m_\pm =m_0 m/|m_0 \pm m|$~\cite{LuttingerSC:th7}. As $\vec{a}_3 (\hat{\Omega},0) \equiv \hat{\vec{d}}(\hat{\Omega})$, besides the trivial $s$-wave pairing ($\Delta_0$), this system supports five even-parity $d$-wave pairings. Each of them hosts two nodal loops on the FS [Fig.~\ref{fig:nodalloopLorentz}(b)]~\cite{LuttingerSC:th7}. The $d$-wave nature of the local pairings stems from the biquadratic normal state band dispersion, also expressed in terms of ``$d$-wave" harmonics. Thus local paired states inherit the normal state band parity, justifying our attribution of the parity mixing or fragmentation in the paired states of RSW materials to its linear normal state band structure.

\begin{figure}[t!]
\includegraphics[width=0.90\linewidth]{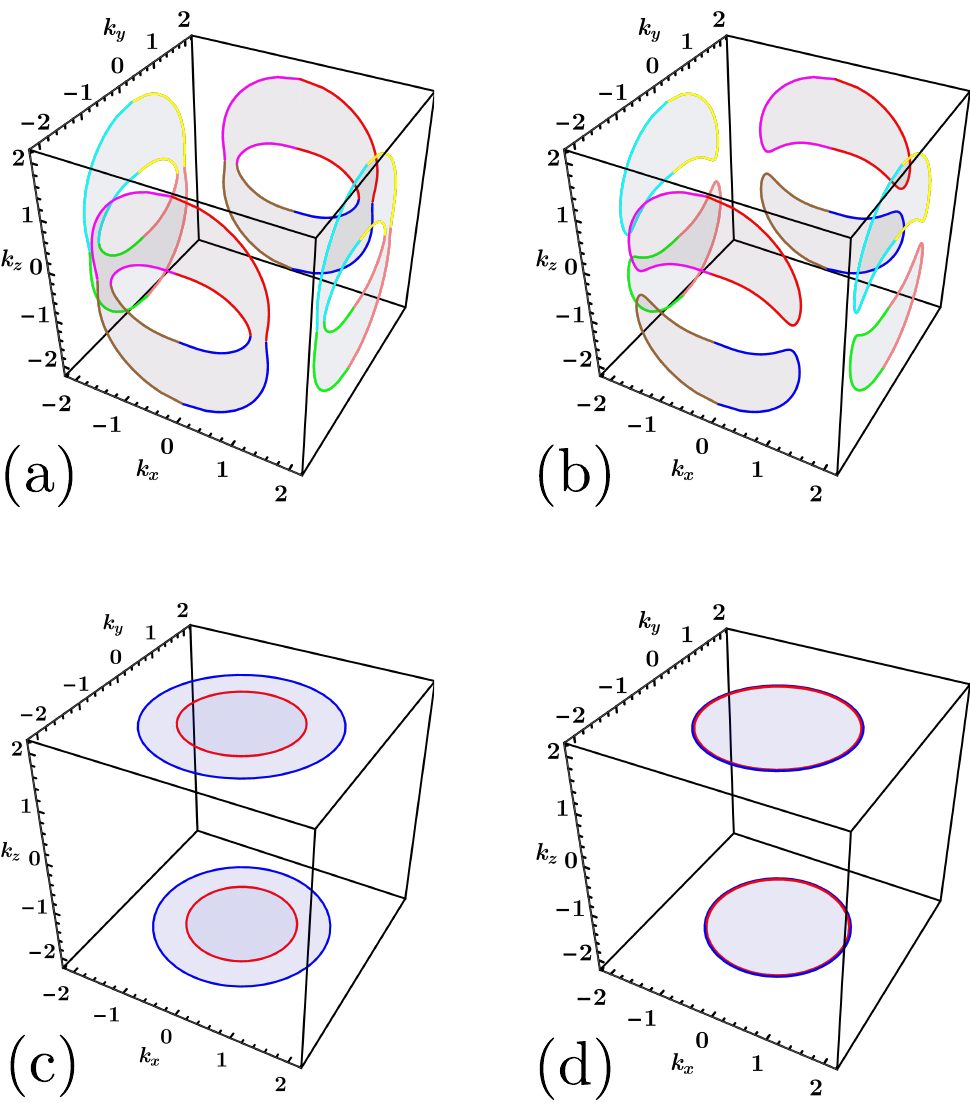}
\caption{Schematic structure of the drumhead flat Majorana surface states (shaded regions) for the $\Delta_3$ (top) and $\Delta_5$ (bottom) pairings, shown in the surface Brillouin zones for (a) $\Delta_3/\mu=0.567$, (b) $\Delta_3/\mu=0.587$, (c) $\Delta_5/\mu=0.628$, and (d) $\Delta_5/\mu=0.707$. Note that the perimeters of the drumhead states are determined by the projections of the bulk polar hairs or the nodal loops, which are shown in Fig.~\ref{fig:nodalloopevolution}, and here color coded accordingly.        
}~\label{fig:drumhead}
\end{figure}

\emph{Gap equation}.~Finally, we compute the pairing amplitudes and transition temperatures ($T_c$) for all non-$s$-wave pairings in a RSW metal within the BCS formalism. We assume that the Debye frequency ($\Omega_D$) over which RSW fermions experience effective attraction is same near two FSs. At finite temperature ($T$), the self-consistent gap equation is obtained by minimizing the free energy 
\allowdisplaybreaks[4]
\begin{equation}~\label{eq:freeenergy}
F_\nu = \frac{\Delta^2_\nu}{2g}-2 \ell^3  k_B T \int \frac{d^3\vec{k}}{(2\pi)^3}\sum_{j=1/2,3/2} \ln \brac{ \cosh \left( \frac{E^\nu_j(\alpha)}{2k_BT} \right)}
\end{equation}
with respect to $\Delta_\nu$. Here $\ell$ is the lattice constant and $g$ is the coupling strength. We numerically solve the gap equation in terms of the dimensionless quantities: $\Delta_\nu/\mu \to \Delta_\nu$, $T/\mu \to t$, $\Omega_D/\mu \to \omega_D$ and $g \ell^3 \mu^2/(2 \pi^2 v^3) \to \lambda$~\cite{supplementary}. Results shown in Fig.~\ref{fig:BCStheory} conform to the universal BCS scaling forms~\cite{BCS:1} 
\allowdisplaybreaks[4]
\begin{equation}~\label{eq:BCSscaling}
\frac{\Delta_\nu}{\omega_D}= a_\nu \exp\left[- \frac{b_\nu}{\lambda}\right],
\frac{k_B t_c}{\omega_D}= c \; \exp \left[ -\frac{b_\nu}{\lambda} \right],
\frac{\Delta_\nu}{k_B t_c}= d_\nu.
\end{equation} 
For $\alpha=0$, $a_j \approx 3.22$ and $d_j=2.84$ for $j=1, \cdots, 4$, $a_5 \approx 3.36$, $d_5 \approx 2.96$, and $b_\nu \approx 0.301$ for all $\nu$. For $\alpha=4/7$, $a_j \approx 2.79 \; (2)$, $b_j=3/2 \; (1/2)$ and $d_j=2.46 \; (1.76)$ for $j=1,2,3$ ($4,5$). But, $c \approx 1.13$ always~\cite{supplementary, footenote:1}.

\emph{Discussions \& outlooks}.~Considering a minimal irreducible four-component continuum model for RSW fermions, here we show that they harbor a plethora of exotic mixed-parity or parity fragmented topological superconductors, featuring polar hairs or nodal rings that reside in between the FSs, showcasing intriguing evolutions with changing pairing amplitude as they then move, collide, and reconnect or disappear. Topological nature of these paired states gives rise to surface-localized drumhead-shaped Majorana Fermi pockets, images of the bulk polar hairs or nodal loops on the surfaces, schematically shown in Fig.~\ref{fig:drumhead}. Polar hairs or nodal rings manifest in specific heat $C_v \sim T^2$~\cite{tinkham}, $T$-linear penetration depth~\cite{prozorov} and inverse  nuclear magnetic resonance relaxation time $1/T_1 \sim T^3$ (Korringa's relation), for example. Due to the $|E|$-linear density of states in the presence of polar hairs or nodal loops, the paired states become a diffusive thermal metal of BdG fermions at lowest temperature even for infinitesimal disorder, yielding $C_v \sim T$ and $1/T_1 \sim T$. Nonetheless, in sufficiently clean systems the aforementioned scaling of physical observables due to polar hairs and nodal loops remains operative over a large temperature window below $T_c$.

In real materials, RSW nodes often appear at finite time-reversal invariant momentum points~\cite{spin32:met1, spin32:met2, spin32:met3, spin32:met4, spin32:met5, spin32:met6, spin32:met7, spin32:met8}, at $\vec{K}$ (say). Then intra RSW node pairings stand as examples of Fulde-Farrell-Larkin-Ovchinikov superconductors for spin-3/2 fermions~\cite{FFLO:1, FFLO:2}, with center of mass momentum $2 \vec{K}$ of Cooper pairs, displaying spatial modulation with periodicity $2 \vec{K}$, as in graphene heterostructures~\cite{FFLO:3, FFLO:4} and spin-1/2 Weyl semimetals~\cite{FFLO:5}. These exciting possibilities should open new avenues in the rich landscape of unconventional superconductors, triggering search for their microscopic Cooper glue (electronic repulsion or phononic attraction) and the role of a material dependent parameter $\alpha$ [Eq.~(\ref{eq:RSWHamil})] in determining the paired state at the lowest temperature in RSW materials. Tabulation of $\alpha$ in different candidate RSW materials will thus be crucial to theoretically predict the nature of the paired states and their competition in these materials. Nonetheless, given that spin-3/2 Luttinger material YPtBi supports nodal pairing~\cite{LuttingerSC:exp4}, it is natural to expect that at least some RSW materials can accommodate the nodal pairings, discussed in this work.

\emph{Acknowledgments}.~S.M. acknowledges support from the KVPY programme. B.R. was supported by NSF CAREER Grant no.\ DMR- 2238679, and thanks Sanjib Kumar Das and Sk Asrap Murshed for discussions.


\begin{thebibliography}{}
\bibitem{footenote:1} In Luttinger materials these coefficients are $a_j \approx 3.335$ and $d_j \approx 2.96$ for $j=1,\cdots, 4$, $a_5 \approx 3.501$, $d_5 \approx 3.09$, $b_\nu =5$ and $c \approx 1.13$ for any $\nu=1,\cdots, 5$~\cite{LuttingerSC:th7}.
\end{thebibliography}
\end{document}